\documentclass[aps,prb,twocolumn,groupedaddress]{revtex4-1}

\usepackage{hyperref}
\hypersetup{
    bookmarks=true,         
    unicode=true,          
    pdftoolbar=false,        
    pdfmenubar=true,        
    pdffitwindow=true,      
    pdftitle={My title},    
    pdfauthor={Author},     
    pdfsubject={Subject},   
    pdfnewwindow=true,      
    pdfkeywords={keywords}, 
    colorlinks=true,       
    linkcolor=red,          
    citecolor=green,        
    filecolor=magenta,      
    urlcolor=cyan           
}

\usepackage[all]{hypcap}
\usepackage{graphicx}
\usepackage{amsmath}

\newcommand{\W}{\Psi^{\rm sc}}
\newcommand{\Wx}{\Psi^{*}}
\newcommand{\w}{\psi^{\rm sc}}
\newcommand{\Wa}{\Phi}
\newcommand{\wa}{\phi}

\newcommand{\R}{{\bf R}}
\newcommand{\dR}{{\bf {\dot R}}}
\newcommand{\Lag}{{\cal L}}

\begin{document}

\title{Wavefunction extended Lagrangian
		Born-Oppenheimer molecular dynamics}

\author{Peter Steneteg$^1$}
\email[]{peter.steneteg@liu.se}
\author{Igor A. Abrikosov$^1$}
\author{Valery Weber$^2$}
\author{Anders M. N. Niklasson$^{3,4}$}
\email[]{amn@lanl.gov}
\affiliation{$^1$Department of Physics, Chemistry and Biology (IFM), 
Link\"oping University, SE-581 83 Link\"oping, Sweden}
\affiliation{$^2$Institute of Physical Chemistry, University of Z\"{u}rich, CH-8057 Z\"{u}rich, Switzerland}
\affiliation{$^3$ Theoretical Division, Los Alamos National Laboratory, Los Alamos, New Mexico 87545, USA}
\affiliation{$^4$ Department of Materials Science and Engineering, Applied Materials Physics, Royal Institute of Technology, SE-100 44 Stockholm, Sweden}

\date{\today}

\begin{abstract}
	Extended Lagrangian Born-Oppenheimer molecular dynamics
    [Niklasson, Phys. Rev. Lett. {\bf 100} 123004 (2008)] has been generalized 
    to the propagation of the electronic wavefunctions. The technique allows highly efficient
    first principles molecular dynamics simulations using plane wave pseudopotential
    electronic structure methods that are stable and energy conserving also under incomplete
    and approximate self-consistency convergence. An implementation of the method within the
    plane-wave basis set is presented and the accuracy and efficiency is
    demonstrated both for semi-conductor and metallic materials. 
\end{abstract}

\maketitle

\section{Introduction}
	As the available computational capacity for scientific computing is growing, first principles 
	Born-Oppenheimer (BO) molecular dynamics (MD) is becoming an increasingly important 
	tool for studying a wide range of material problems. 
	First principles BOMD delivers a very accurate approach to atomistic simulations 
	without relying on a fitted parameterization of the atomic interactions as in classical 
	molecular dynamics. 
    Unfortunately, applications of BOMD that are based on self-consistent field (SCF)
    calculations such as density functional theory \cite{Hohenberg:1964p2861,Kohn:1965p2886,DMarx00}
    are often limited by a very high computational cost or by fundamental
    shortcomings such as unbalanced phase space trajectories,
    numerical instabilities and a systematic long-term energy drift 
    \cite{PPulay04,Herbert:2005p4322,Niklasson:2006p66}. 
    
    Recently an extended Lagrangian BOMD (XL-BOMD) was 
    introduced that avoids some of the most serious problems of regular BOMD 
    and enables computationally efficient and stable simulations of energy conserving 
    (microcanoncial) ensembles \cite{Niklasson:2008p3161,Niklasson:2009p3195,AOdell09,Niklasson:2006p66}. 
    In XL-BOMD, auxiliary electronic degrees of freedom are included, in addition to the nuclear coordinates and velocities. 
    In contrast to the popular extended Lagrangian Car-Parrinello molecular dynamics
    methods \cite{DMarx00,Car:1985p4316,HBSchlegel01,JMHerbert04}, 
    the nuclear forces are calculated at the ground state BO potential energy 
    surface and the total BO energy is a constant of motion. 

    So far XL-BOMD has been limited to density matrix formulations of the extended
    electronic degrees of freedom. 
    This excludes any practical implementation in widely used plane-wave pseudopotential schemes, 
    since it would lead to unmanageable large density matrices. 
    Because of the arbitrary phase  of the electronic wavefunctions 
    \cite{Arias:1992p1418,Kresse:1993p4205,Kresse:1994p4077} it is difficult to use wavefunctions as the extended 
    electronic degrees of freedom in a stable time-reversible or geometric integration of the equations of motion.
    Here we show how the electronic wavefunctions can be included in XL-BOMD.
    Our formulation allows a time-reversible integration of both the nuclear and the 
    electronic degrees of freedom and it provides a highly efficient BOMD 
    for plane-wave pseudopotential methods that is stable and energy conserving also
    under incomplete and approximate SCF convergence. The wavefunction XL-BOMD method was implemented in the 
    Vienna Ab-initio Simulation Package (VASP) 
    \cite{Kresse:1996p4007,Kresse:1996p4006,Kresse:1993p4005} and its accuracy and efficiency are
    demonstrated both for semi-conductor and metallic materials.

\section{First principle molecular dynamics}	
\subsection{Born-Oppenheimer molecular dynamics}
	First principle BOMD based on density functional theory (DFT)
	is given by the Lagrangian,
	\begin{equation}
		\Lag^{BO}(\R,{\dR}) = 
		\frac{1}{2}\sum_i M_i \dot{R}^2_i - U_{\rm{DFT}}[\R;\W], 
	\end{equation}
	where $\R=\{R_i\}$ are the nuclear coordinates and the dot denotes the time derivative. 
	The potential $U_{\rm{DFT}}[\R;\W]$ is the ground state energy, including ion-ion repulsions, 
        for the density given by the self consistent (sc)
	electronic wavefunctions, $\W = \{\w_{nk}\}$.  
    Here $n$ and $k$ denote the band and reciprocal lattice vectors, respectively. 
    The Euler-Lagrange equations,
	\begin{equation}
		\frac{d}{dt}\left(\frac{\partial\Lag}{\partial \dot{R_i}}\right) - \frac{\partial\Lag}{\partial R_i}  = 0
	\end{equation}
	give the equations of motion for the dynamical variables $\R(t)$ and $\dR(t)$.

    The high cost of finding the ground state SCF solution $\W(t)$ is significantly reduced 
    by using an initial guess that is extrapolated from previous time steps 
    \cite{Arias:1992p1077,JMMillan99,DAlfe99,PPulay04,CRaynaud04,Herbert:2005p4322},
	\begin{equation}
		\label{eqn:SCFOPT_0}
		\W(t) = {\rm SCF}[\sum_{m = 1}^M c_m \W (t-m\delta t);\R]. 
	\end{equation}
    In the SCF optimization in Eq.\ (\ref{eqn:SCFOPT_0}) above we assume a full optimization, which may include several iterative 
    cycles based on, for example, simple linear mixing, Broyden mixing, or the direct inversion of the iterative subspace (DIIS) method 
    \cite{PHDederichs83,DDJohnson88,GKarlstrom79,PPulay80}. However, since the SCF optimization in practice never is complete, 
    the extrapolation procedure in Eq.\ (\ref{eqn:SCFOPT_0}) leads to an irreversible evolution of the ground state electronic wavefunctions. 
    The nuclear forces are therefore calculated with an underlying electronic degrees of freedom that 
    behave unphysically. 
    The irreversibility appears most strikingly as a systematic long-term energy 
    drift\cite{PPulay04,Herbert:2005p4322}. 
    By using thermostats, e.g. an artificial interaction with an external heat bath, these shortcomings of
    BOMD may not be noticed. However, a thermostat requires an underlying dynamics that is physically correct, 
    and the problems are therefore never removed.  Only by improving the SCF convergence, which is increasing the 
    computational cost, is it possible to suppress the energy drift, though the problem 
    never fully disappears.

\subsection{Wavefunction Extended-Lagrangian Born-Oppenheimer MD}
	In our  wavefunction XL-BOMD, proposed here, the dynamical variables of the  
	BO Lagrangian are extended with a set of auxiliary wavefunctions $\Wa = \{\wa_{nk}\}$ evolving in
	harmonic oscillators centered around the self-consistent ground state wavefunctions $\W(t)$, 
	\begin{align}
		\label{eqn:lxbo}
		\Lag^{\rm XBO}(\R,{\dR}, \Wa, \dot{\Wa} ) 
		 =    
		\Lag^{\rm BO} + 
		\frac{\mu}{2}\sum_{nk}\int |\dot{\wa}_{nk}|^2 d{\bf r} &  \nonumber \\
		-  \frac{\mu \omega^2}{2}\sum_{nk}\int|\w_{nk}-\wa_{nk}|^2 d{\bf r}.&
	\end{align}
	Here $\mu$ is a fictitious electron mass parameter and $\omega$ is a 
	frequency or curvature parameter for the harmonic potentials.
	Applying the Euler-Lagrange equations to the extended Lagrangian in Eq.\ (\ref{eqn:lxbo}) gives
	\begin{equation}
		M_i\ddot{R_i} =  
		-  \frac{\partial U_{\rm DFT}}{\partial R_i}	
		-  \frac{\mu \omega^2}{2}\frac{\partial}{\partial R_i} \sum_{nk}\int|\w_{nk}-\wa_{nk}|^2 d{\bf r},
	\end{equation}	
	\begin{equation}
		\mu\ddot{\Wa}(t) = \mu\omega^2\big(\W(t)-\Wa(t)\big).\\
	\end{equation}
	In the limit $\mu\rightarrow 0$ we get 
	\begin{equation}
		\label{eqn:BOMDEM}
		M_k\ddot{R_i} = - \frac{\partial U_{\rm DFT}[\R;\W]}{\partial R_i},
	\end{equation}
	\begin{equation}
		\label{eqn:XLEM}
		\ddot{\Wa}(t) = \omega^2\big(\W(t)-\Wa(t)\big).\\
	\end{equation}
	Thus, in the limit of vanishing fictitious mass parameter, $\mu$, we recover 
        the regular BO equations of motion in Eq.\ (\ref{eqn:BOMDEM}), with the total BO energy as a constant of motion.  
        Equation (\ref{eqn:XLEM}) determines the dynamics of our auxiliary wavefunctions $\Wa(t)$. 
        Since $\mu$ is set to zero, the only remaining undetermined parameter is the frequency 
        or curvature $\omega$ of the extended harmonic potentials.
        As will be shown below, $\omega$ occurs in the integration of Eq.\ (\ref{eqn:XLEM}) only as a dimensionless factor 
        $\delta t^2 \omega^2$ and therefore affects the dynamics in the same way as the finite integration time step $\delta t$.

    Since the auxiliary wavefunctions $\Wa(t)$ are dynamical variables, they can 
    be integrated by, for example, the time-reversible Verlet algorithm \cite{Verlet:1967p4317}.
	Morevover, since the auxiliary wavefunctions evolve in a harmonic well centered around the ground state solution, 
    $\Wa(t)$ will stay close $\W(t)$. By maximizing the curvature $\omega^2$ of the harmonic extensions we can minimize their 
    separation. Using the auxiliary dynamical variables $\Wa(t)$ in the initial guess to the SCF optimization,
	\begin{equation}
		\label{eqn:SCFOPT}
		\W(t) = {\rm SCF}\left[\Wa(t);\R\right],
	\end{equation}
    therefore provides an efficient SCF procedure that can be used within a time-reversible framework.
    The nuclear forces will then be calculated with an underlying electronic degrees
    of freedom with the correct physical time-reversal symmetry. This is in contrast to
    conventional BOMD, where the SCF optimization is given
    from an irreversible propagation of the underlying electronic degrees of freedom as in Eq.\ (\ref{eqn:SCFOPT_0}).
	Hence the system will be propagated reversibly and should not suffer from any systematic 
	drift in the total energy and phase space.
	
\subsection{Integration}
	Both the nuclear and electronic degrees of freedom in Eqs.\ (\ref{eqn:BOMDEM}) and 
	(\ref{eqn:XLEM}) can be integrated with the Verlet algorithm, or with other
   	geometric integration schemes that preserve properties of the exact underlying flow
        of the dynamics\cite{BLeimkuhler04,Niklasson:2008p3161,AOdell09}.
	The Verlet integration of Eq.\ (\ref{eqn:XLEM}), including a weak external dissipative electronic 
	force that removes accumulation of numerical noise \cite{Niklasson:2009p3195}, has the following form
	\begin{align}
		\label{eqn:verlet}
		\Wa(t+\delta t) = 2\Wa(t) - \Wa(t -\delta t) &  \nonumber \\ + \delta t^2 \omega^2\big(\W(t)-\Wa(t)\big) 
		 +	\alpha \sum_{m=0}^K c_m \Wa(t-m\delta t).
	\end{align}	
	where $\alpha$ determines the magnitude of the dissipative force term with the $c_m$ 
	coefficients given in Ref. \cite{Niklasson:2009p3195}.
	The additional electronic force introduces dissipation of 
    numerical noise that would accumulate in a perfectly reversible and lossless propagation. The dissipation
    breaks time-reversibility, but only to a high order in $\delta t$ 
    \cite{Niklasson:2009p3195,Kolafa:2004p4320}.
    In this way numerical errors can be removed without causing 
    any significant drift in the total energy. 

\begin{table}[t]
  \centering
  \caption{\protect Coefficients for the Verlet integration scheme with the external dissipative
  force term in Eq.\ (\ref{eqn:verletU}). The coefficients are derived in Ref.\ \cite{Niklasson:2009p3195}, which
  contains a more complete set of coefficients.
  }\label{Tab_Coef}
  \begin{ruledtabular}
  \begin{tabular}{llccccccccc}
    $K$ & $\delta t^2 \omega^2$ & \!\!$\alpha\!\times\!10^{-3}$\!\! & \!$c_{0}$ & $c_{1}$ & $c_{2}$ & $c_{3}$ & $c_{4}$ &
$c_{5}$ & $c_{6}$ & $c_{7}$ \\
     \hline
     0  & 2.00 & \!\!0\!\! &   &    &   &  &  &      &     &     \\
     3  & 1.69 & \!\!150\!\! & -2    &   3   &   0  & -1  &     &      &     &     \\
     5  & 1.82 & \!\!18\!\! &-6    &   14   &  -8  & -3  &  4  &  -1  &     &      \\
     7  & 1.86 & \!\!1.6\!\! &-36    &   99  & -88  & 11  &  32 & -25  & 8   &  -1  \\
  \end{tabular}
  \end{ruledtabular}
\end{table}

	\subsection{Subspace alignment}
    The electronic ground state wavefunctions are unique except with respect to their phase.
    This presents a problem for the accuracy and stability of the Verlet integration above. The 
    wavefunctions need to be aligned to a common orientation to allow an accurate and stable integration. 
    Aligning the wavefunctions backwards in time as in some previous integration schemes for regular BOMD
    \cite{Arias:1992p1418,PGiannozzi09,Kresse:1994p4077,Kresse:1993p4205} is 
    not possible, since it would break the time-reversal symmetry.
    We solve this problem by including a unitary rotation transform $U$ in the SCF optimization,
    which rotates $\W(t)$ such that the deviation from $\Wa(t)$ is minimized in the Frobenius norm, i.e.
	\begin{equation}
		U = \mbox{arg} \min_{U'} || \W(t) U' - \Wa(t) ||_F.
	\end{equation}	
	$U$ can be calculated from $U = (O O^{\dagger})^{-1/2} O$ 
	%
	%
	where $O = \langle \W | \Wa \rangle$ is the overlap matrix between $\W(t)$ and $\Wa(t)$ \cite{Arias:1992p1418,Kresse:1993p4205,PGiannozzi09}. 
	Since the rotation is only applied to $\W(t)$ and not to previous auxiliary wavefunctions, 
	the reversibility is not affected. The redefined Verlet integration is
	\begin{align}\label{eqn:verletU}
		\Wa(t+\delta t) = 2\Wa(t) - \Wa(t -\delta t)  &  \nonumber \\ + \delta t^2 \omega^2\big( \W(t)U-\Wa(t) \big) 
                +      \alpha \sum_{m=0}^K c_m \Wa(t-m\delta t).
	\end{align}	
     Note that good initial values for the axillary variables are important. If a poor initial guess are
     used the weak dissipation will eventually relax the auxiliary dynamics to a similar dynamics, but it would 
     take time, and meanwhile we would have bad initial guesses for the SCF optimization. In our 
     implementation the initial values of the auxiliary variables, are set to the SCF optimized 
     ground states, i.e. as intial conditions for $\Wa(t)$ we chose to set $\Wa(t) \equiv \W(t)$ for the first $K+1$ time steps,
     where we perform phase alignements to the first optimized wavefunctions $\W(t = t_0)$.   
     It may be preferable to run with a stronger convergence criteria during the initial 
     steps to get a good starting guess. In the first $K+1$ initial steps we therefore chose 
     to have a higher degree of SCF convergence than in later time steps.

	\section{Stability and noise dissipation}
\begin{figure}[htb]
 		\centering
  		\includegraphics[scale=1]{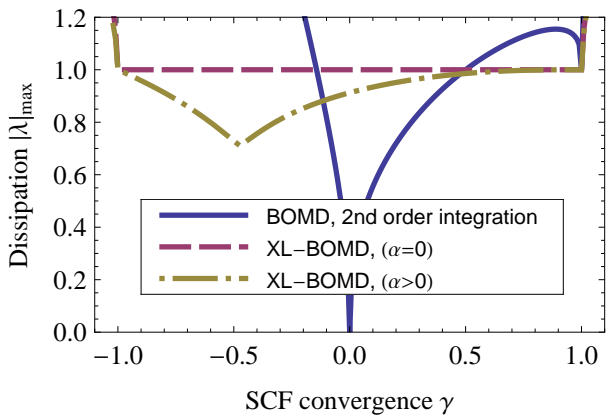}
  		\caption{ 
  			(Color online) Stability and dissipation for 2nd order regular BOMD\cite{Arias:1992p1418,Kresse:1993p4205}
                        and XL-BOMD, Eq.\ (\ref{eqn:verletU}). The stability region of the regular BOMD
                        is limited to $\gamma\in[-0.14, 0.50]$ hence demanding a 
                        higher degree of SCF convergence, even if the accuracy in each step is high.
                        In contrast, XL-BOMD is stable in the entire region of SCF convergence, $\gamma\in[-1, 1]$.}
 		\label{fig:Stability}
	\end{figure}

    By aligning the phase in the SCF optimization, the stability of the Verlet integration in Eq.\ (\ref{eqn:verletU}),
    under the condition of an approximate and incomplete SCF convergence, can be analyzed from 
    the roots $\lambda$ of the characteristic equation of the homogeneous (steady state) part of the Verlet scheme, in the
    same way as for the density matrix \cite{Niklasson:2008p3161,Niklasson:2009p3195}. 
    Assume a linearization of an approximate SCF
    optimization, Eq.\ (\ref{eqn:SCFOPT}), around the hypothetical exact solution ${\Wx}$, where
    \begin{equation}\label{LinSCF}
    \W = {\rm SCF}[\Wa] \approx {\Wx} + \Gamma_{\rm SCF} (\Wa-{\Wx}).
    \end{equation}
    Let $\gamma$ be the largest eigenvalue of the SCF response kernel $\Gamma_{\rm SCF}$.
    Inserting Eq.\ (\ref{LinSCF}) in the Verlet scheme, Eq.\ (\ref{eqn:verletU}), with $\Gamma_{\rm SCF}$ replaced
    by $\gamma$, for the homogeneous steady state solution for which ${\Wx} \equiv 0$, gives the characteristic equation
    \begin{align}
    	\lambda^{n+1} = 2\lambda^n -\lambda^{n-1} + \kappa(\gamma-1)\lambda^n  &  \nonumber \\
                + \alpha \sum_{m=0}^K c_m \lambda^{n-m}.
    \end{align}
   Here the dimensionless constant $\kappa = \delta t^2 \omega^2$, and $\gamma \in [-1,1]$ 
   is proportional to the amount of convergence in the SCF optimization.
   As long as the initial guess $\Wa(t)$ is brought closer to the ground state solution by the SCF procedure,
   $|\gamma|$ will be smaller than 1. If the characteristic roots have
   a magnitude $|\lambda|_{max}>1$, the integration is unstable (even if the accuracy is good), 
   whereas it is stable if $|\lambda|_{max}\leq 1$ (even if the optimization is approximate).
   For $|\lambda|_{max} < 1$ the accumulation of numerical noise will be suppressed through dissipation.
   By optimizing $\kappa = \delta t^2 \omega^2$ under the condition of stability under incomplete SCF convergence with $\gamma \in [-1,1]$, 
   the curvature $\omega^2$ of the extended harmonic wells will be maximized, which keeps the auxiliary wavefunctions $\Wa(t)$ as close
   as possible to the ground state solutions $\W(t)$ \cite{AOdell09}. This optimization is performed under the 
   additional condition of maximum dissipation.  Our optimized values of $\alpha$ and $\kappa$ 
   and the $c_m$ coefficients can be found in Ref. \cite{Niklasson:2009p3195} and a few examples are given in Tab.\ \ref{Tab_Coef}.
   Three different examples of dissipation as a function of SCF convergence as measured by $|\lambda|_{\rm max}$ 
   and $|\gamma|$ are shown in Fig.\ \ref{fig:Stability}. 
	%

	\section{Plane wave pseudo-potential implementation (VASP)}	
	Our wavefunction XL-BOMD method has 
	been implemented in the Vienna Ab-initio Simulation 
	Package (VASP) \cite{Kresse:1996p4007,Kresse:1996p4006,Kresse:1993p4005}
	and the projector augmented wave method \cite{Blochl:1994p1407,Kresse:1999p4180}.
	These particular methods not only require the integration of the wavefunctions, 
	but also the electron density and the Kohn-Sham eigenvalues that are used in the SCF optimization.
    In this work these additional quantities has been be added to the Lagrangian as extended
    dynamical variables evolving in harmonic oscillators centered around their own optimized values
    in the same way as the auxiliary wavefunctions. For example, an auxiliary density, $\rho({\bf r})$, can be included as a dynamical variable
    through the extended Lagrangian,
        \begin{align}
                \label{eqn:lxbo_rho}
                \Lag^{\rm XBO'}(\R,{\dR}, \Wa, \dot{\Wa},\rho,\dot{\rho} )
                 = \Lag^{\rm XBO}(\R,{\dR}, \Wa, \dot{\Wa} ) & \nonumber \\
      + \frac{1}{2} \mu\int {\dot \rho({\bf r})}^2 d{\bf r} - \frac{1}{2} \mu \omega^2 \int \left(n^{\rm sc}({\bf r}) - \rho({\bf r}) \right)^2d{\bf r} .
        \end{align}
     Here $\rho({\bf r})$ follows the SCF optimized ground state density $n^{\rm sc}({\bf r})$. Also for the auxiliary density
     and other extended variables the initial values are set equal to the optimized ground state values for the first $K+1$ steps.

%
	\begin{figure}[htb]
 		\centering
  		\includegraphics[scale=1]{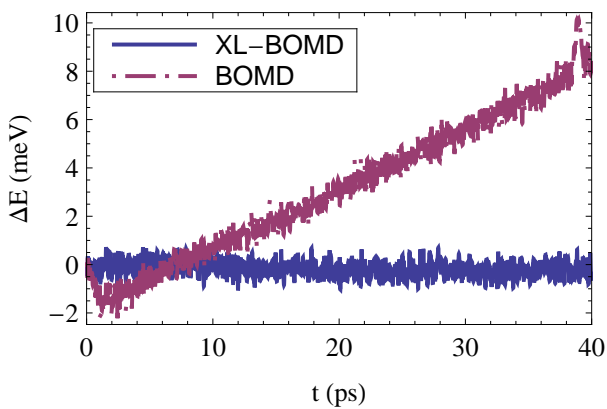} 
  		\caption{(Color online) Fluctuations in the total energy, $\Delta E$, as a function of time for a Na 
  			bcc crystal with 16 atoms in the unit cell and an integration time step of 4 fs. The same
                        SCF convergence criterion was used, $\delta E =  5$ $\mu$eV, requiering about 2 SCF iterations 
                        per time step for both methods.  The regular BOMD simulation shows significant systematic energy drift. }
  		\label{fig:Na-Etot}
	\end{figure}
\section{Applications}
	To demonstrate the accuracy and efficiency of the wavefunction XL-BOMD scheme
    we simulate two different model systems with qualitatively different bonding, 
    metallic sodium and semiconducting silicon. 
\subsection{Sodium}
	A unit cell of 16 bcc Na atoms was simulated for a total of 10,000 steps
	with an ionic temperature fluctuating around 500 K with a time step of 4 fs. 
	The regular BOMD integration scheme was based on a 2nd-order extrapolation of the
	wavefunctions from three previous time steps \cite{Arias:1992p1418,Kresse:1993p4205} and the XL-BOMD scheme used
    $K=5$ for the dissipation \cite{Niklasson:2009p3195}. Both methods 
    used the velocity Verlet integration for the nuclear degrees of freedom and
	were run with the same SCF energy convergence criterion, $\delta E =  5$ $\mu$eV, 
	resulting in about 2 SCF iterations per time step for both methods.  Each SCF cycle includes
	one single construction and solution of the Hamiltonian eigenvalue problem. 
	As SCF convergence accelerating algorithm we used the DIIS scheme \cite{PPulay80}.
	A plane-wave energy cutoff of 102 eV and a grid of 64 k-points was used and the 
	exchange-correlation energy was given by the local density approximation (LDA)\cite{Ceperley:1980p4515}.

	The fluctuations in the total energy can be seen in Fig. \ref{fig:Na-Etot}. For regular BOMD we 
	see a small but systematic drift in the total energy of the order of 0.25 meV/ps. 
	In comparison, XL-BOMD shows no drift and the magnitude of the energy fluctuations
	due to the local truncation errors, occurring because of the finite time steps and the approximate SCF convergence, is the same. 
    In fact, we have found that XL-BOMD is stable even when only 1 SCF cycle per time step is used. 
	This would be a general statement if the SCF procedure systematically improves the convergence 
	in a single step\cite{Niklasson:2008p3161}. Unfortunately, this is not always the case.

\subsection{Silicon}
	Next a Si system with 8 atoms per unit cell in a diamond structure was simulated 
	for a total of 10,000 steps at an ionic temperature fluctuating around 500 K with a time step of 1 fs. 
	The SCF convergence threshold $\delta E$ was set to 5 $\mu$eV with a plane wave cutoff of 246 
	eV and a grid of 64 k-points was used. Otherwise the same settings as for sodium were applied.
\begin{figure}[htb]
  \centering
  \includegraphics[scale=1]{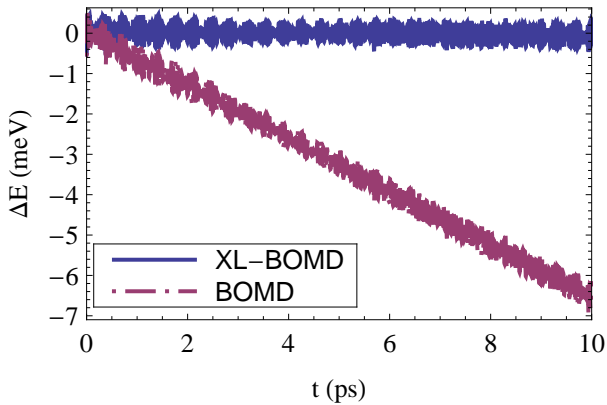} 
  \caption{(Color online) Fluctuations in the total energy, $\Delta E$ versus time for 8 Si atoms
  		simulated using XL-BOMD and regular BOMD. 
   		 BOMD shows a systematic energy drift.}
  \label{fig:Si-Etot}
\end{figure}
%
	In Fig. \ref{fig:Si-Etot} the fluctuations in the total energy $\Delta E$ is plotted. 
	Also in this case, we find that XL-BOMD restores balance to the unphysical trajectories of regular BOMD
        that shows a significant systematic drift in the total energy.

\section{Discussion and Summary}
   	In a direct comparison using the same time step and convergence criteria, we find that XL-BOMD 
    and regular BOMD have the same local truncation error, as measured by the local amplitude of the oscillations
    in the total energy, both for the metallic and the non-metallic system. 
    However, the unphysical behavior of regular BOMD, which has a  systematic long-term energy drift, 
    is removed in XL-BOMD. Only by significantly increasing
    the computational cost of regular BOMD with a higher degree of SCF convergence, or shorter time steps, 
    is it possible to reduce the long-term energy drift.
    Table \ref{tbl:efficiency} summarizes the results of a comparison between XL-BOMD and regular BOMD for the Na simulation. 
    The results clearly show that even though the energy drift can be substantially reduced in conventional 
    BOMD, a large performance penalty has to be paid. 
    XL-BOMD requires in general more memory, 1.5 to 2.5 times the temporary storage used in regular BOMD depending on the dissipation scheme used. 
    However, for most practical situations wall time is the limiting factor when running first principal BOMD, not memory usage.
    XL-BOMD therefore combines a more correct physical description with a lower computational cost. 

    In many ways, XL-BOMD integrates some of the best features of regular BOMD and Car-Parrinello molecular dynamics, i.e 
    the parameter-free rigor of BOMD and an efficient extended Lagrangian framework as in Car-Parrinello molecular dynamics, 
    where both nuclear and electronic degrees of freedom are included as dynamical variables.
	\begin{table}
	\caption{\label{tbl:efficiency} Comparison between
		regular BOMD and XL-BOMD for a 16 atom Na bcc simulation. 
                 Time step, $\delta t$, in fs, energy convergence threshold, $\delta E$, in $\mu$eV, number of SCF cycles, and systematic 
                 Drift (per atom) in $\mu$eV/ps. Drift $<$ 0.1 means no systematic drift was found.}
	\begin{ruledtabular}
	\begin{tabular}{lccccc}
 		\text{Method} & \text{$\delta t$} & \text{$\delta E$} & \text{SCF} & \text{SCF/fs} & \text{Drift} \\
 		\hline
 		\text{XL-BOMD} 				& 4 & $5.0$ & 2.04 & 0.51 & $<$ 0.1 \\
 		\text{XL-BOMD 1 SCF} 	& 4 & - & 1.00 & 0.25 & $<$ 0.1 \\
 		\text{BOMD} 					& 4 & $5.0$ & 2.39 & 0.6  & 15.6 \\
 		\text{BOMD short step} 	& 1 & $5.0$ & 2.06 & 2.05 & 0.8 \\
 		\text{BOMD high conv.}	& 4 & $5\!\cdot\!10^{-4}$ & 4.72 & 1.18 & 1.8 
	\end{tabular}
	\end{ruledtabular}
	\end{table}

	In summary, we have proposed and demonstrated a wavefunction XL-BOMD 
    scheme that allows highly efficient first principles
    molecular dynamics simulations using plane wave pseudopotential electronic structure 
    methods that are stable and energy conserving also under incomplete and approximate 
    self-consistency convergence.   This extends the capability and accuracy
    of modern molecular dynamics simulations.
	
	\begin{acknowledgments}
	We gratefully acknowledge the support of the Swedish Foundation for Strategic Research (SSF) via
	the Strategic Research Center MS$^2$E, the G\"oran Gustafsson Foundation for Research in Natural
	Sciences and Medicine, and the US Department of Energy through the LANL LDRD/ER program 
	for this work, as well as Nicolas Bock and Travis Peery for the stimulating environment at the T-Division Ten-Bar Java group.
	\end{acknowledgments}

%

\end{document}